\documentclass[sigconf]{acmart}

\copyrightyear{2022}
\acmYear{2022}
\setcopyright{rightsretained}
\acmConference[ICSE '22 Companion]{44th International Conference on
Software Engineering Companion}{May 21--29, 2022}{Pittsburgh, PA, USA}
\acmBooktitle{44th International Conference on Software Engineering
Companion (ICSE '22 Companion), May 21--29, 2022, Pittsburgh, PA, USA}
\acmDOI{10.1145/3510454.3516851}
\acmISBN{978-1-4503-9223-5/22/05}

\usepackage[utf8]{inputenc}
\usepackage[inline]{enumitem} 
\usepackage{soul}
\usepackage{xcolor}
\usepackage{xspace}
\usepackage{balance}

\usepackage{graphicx}
\usepackage{listings,multicol}
\usepackage[caption=false]{subfig}  
\usepackage[export]{adjustbox}
\usepackage{amsmath}
\usepackage{textcomp}
\usepackage{listings}
\usepackage{cleveref}
\usepackage{algorithm}
\usepackage{float}
\usepackage[noend]{algpseudocode}

\newcommand{\ie}{\emph{i.e.,}\xspace}
\newcommand{\eg}{\emph{e.g.,}\xspace}

\newboolean{showcomments} 
\setboolean{showcomments}{true}
\ifthenelse{\boolean{showcomments}}
{\newcommand{\nb}[2]{
		\fbox{\bfseries\sffamily\scriptsize#1}
		{\sf\small$\blacktriangleright$\textit{\textcolor{red}{#2}}$\blacktriangleleft$}
	}
}
{\newcommand{\nb}[2]{}}

\ifthenelse{\boolean{showcomments}}
{
}
{}

\newcommand{\toolname}{\textsc{IntelliTC}\xspace}
\newcommand{\tcinfer}{\textsc{TC-Infer}\xspace}

\newcommand{\code}[1]{\texttt{\fontsize{8.5}{10}\selectfont #1}}
\newcommand{\smcode}[1]{\texttt{\fontsize{7.5}{8}\selectfont #1}}

\colorlet{punct}{red!60!black}
\definecolor{background}{HTML}{EEEEEE}
\definecolor{delim}{RGB}{20,105,176}
\colorlet{numb}{magenta!60!black}

\lstdefinelanguage{json}{
    basicstyle=\scriptsize\ttfamily,
    numbers=left,
    numberstyle=\scriptsize,
    stepnumber=1,
    numbersep=8pt,
    showstringspaces=false,
    breaklines=true,
    frame=lines,
    backgroundcolor=\color{background},
    literate=
     *{0}{{{\color{numb}0}}}{1}
      {1}{{{\color{numb}1}}}{1}
      {2}{{{\color{numb}2}}}{1}
      {3}{{{\color{numb}3}}}{1}
      {4}{{{\color{numb}4}}}{1}
      {5}{{{\color{numb}5}}}{1}
      {6}{{{\color{numb}6}}}{1}
      {7}{{{\color{numb}7}}}{1}
      {8}{{{\color{numb}8}}}{1}
      {9}{{{\color{numb}9}}}{1}
      {:}{{{\color{punct}{:}}}}{1}
      {,}{{{\color{punct}{,}}}}{1}
      {\{}{{{\color{delim}{\{}}}}{1}
      {\}}{{{\color{delim}{\}}}}}{1}
      {[}{{{\color{delim}{[}}}}{1}
      {]}{{{\color{delim}{]}}}}{1},
}

\title{\toolname: Automating Type Changes in IntelliJ IDEA}

\begin{document}

\author{Oleg Smirnov}
\affiliation{%
    \institution{JetBrains Research}
    \country{Republic of Serbia}
}
\email{oleg.smirnov@jetbrains.com}

\author{Ameya Ketkar}
\authornote{Ameya Ketkar performed this work as part of his PhD at Oregon State University.}
\affiliation{%
  \institution{Uber Technologies Inc.}
  \country{USA}
}
\email{ketkara@uber.com}

\author{Timofey Bryksin}
\affiliation{%
    \institution{JetBrains Research}
    \country{Republic of Cyprus}
}
\email{timofey.bryksin@jetbrains.com}

\author{Nikolaos Tsantalis}
\affiliation{%
  \institution{Concordia University}
  \country{Canada}
}
\email{nikolaos.tsantalis@concordia.ca}

\author{Danny Dig}
\affiliation{%
  \institution{University of Colorado Boulder}
  \country{USA}
}
\email{danny.dig@colorado.edu}

\begin{abstract}
Developers often change types of program elements.
Such refactoring often involves updating not only the type of the element itself, but also the API of all type-dependent references in the code, thus it is tedious and time-consuming.
Despite type changes being more frequent than renamings, just a few current IDE tools provide partially-automated support only for a small set of hard-coded types.
Researchers have recently proposed a data-driven approach to inferring API rewrite rules for type change patterns in Java using code commits history.
In this paper, we build upon these recent advances and introduce \toolname~---~a tool to perform Java type change refactoring.
We implemented it as a plugin for IntelliJ IDEA, a popular Java IDE developed by JetBrains.
We present 3 different ways of providing support for such a refactoring from the standpoint of the user experience: Classic mode, Suggested Refactoring, and Inspection mode.
To evaluate these modalities of using \toolname, we surveyed 22 experienced software developers. 
They positively rated the usefulness of the tool.

The source code and distribution of the plugin are available on GitHub: \url{https://github.com/JetBrains-Research/data-driven-type-migration}.
A demonstration video is available on YouTube: \url{https://youtu.be/pdcfvADA1PY}.
\end{abstract}

\maketitle

\section{Introduction}\label{sec:introduction}

As the program evolves, developers change the type of variables and methods for several reasons, such as library migration (\eg \smcode{org.apache.commons.logging.Log} to \smcode{org.slf4j.Logger}), security (\eg \smcode{java.util.Random} to \smcode{java.security.SecureRandom}), or performance (\smcode{StringBuffer} to \smcode{StringBuilder}).
From a developer's perspective, such \textit{type change} refactoring is much more complicated and tedious than just changing the type of some identifier.
To perform a type change, developers update the declared type of a program element (\eg local variable, parameter, field, or method) and adapt the code referring to this element (within its lexical scope) to the API of the new type.
Due to assignments, argument passing, or public field access, a developer might perform a \textit{series} of type changes to propagate type constraints for the new type.

In our empirical study investigating the practice of type changes in the real world, \citet{ketkar2020understanding} observed that type changes are performed more often than popular refactorings like \textit{Rename}.
In contrast to other refactoring types that are heavily automated by all popular integrated development environments (IDEs), no IDE actively automates type changes; thus developers have to perform most of them manually. 
The state-of-the-practice type change automation tool provided by IntelliJ IDEA~\cite{IDEA} is only applicable to a small set (around ten) of hard-coded type changes that eliminate the use of deprecated types from the Guava library or pre-Java 8 APIs, and it does not allow developers to express and adapt \textit{custom} type changes. 
While state-of-the-art type migration tools \cite{balaban2005refactoring,ketkar2019type,wright2020incremental} are more applicable from the aspect of allowing users to automate custom type changes, these tools are either not supported or are distributed as stand-alone applications depending upon specific static analysis frameworks like Google's Error Prone~\cite{error-prone} or clang project's \textsc{LibTooling} infrastructure~\cite{clang}.
This greatly limits their applicability and usefulness in practice, because 
\begin{enumerate*}
    \item using these external tools breaks developer workflows while working in an IDE~\cite{johnson2001you},
    \item not all developers use (or can use) these specific static analysis frameworks, and  
    \item the user has to \textit{handcraft} the transformation specifications required to perform the custom type change
\end{enumerate*}.

In this paper, we introduce \toolname, a developer-friendly IntelliJ IDEA plugin for automating type changes, and explore its UI/UX aspects.
\toolname leverages the underlying IntelliJ's Type Migration framework~\cite{typeMigration} to allow developers to express type changes as rewrite rules over Java expressions using IntelliJ's Structural-Search-and-Replace templates~\cite{SSR}.
\toolname provides three modes to automate type changes in multiple developer workflows.
For instance, in the \textit{Classic} mode \toolname has to be manually invoked (similar to \textit{Rename} refactoring), while in the \textit{Inspection} mode \toolname recommends type changes.

In our accompanying paper~\cite{ketkar2022tcinfer}, we describe \tcinfer~---~a tool that automatically infers API rewrite rules required to perform type changes, by analyzing the version history of Java projects.
This reduces the burden on the developers, since they no longer have to handcraft the rules for popular type changes. 
We identified 59 most popular and useful type changes (and the associated rewrite rules) reported by \tcinfer. 
We then applied \toolname (a plugin we developed in this work) to replicate 3,060 instances of these 59 type changes encountered in commit histories. 
Our results showed that \toolname has 99.2\% accuracy at automating type changes. 

We have also surveyed 22 experienced software developers to evaluate the potential usefulness of different features that \toolname offers. 
The participants have positively rated the chosen ways of UI envisioning, and confirmed the usefulness of the idea of employing popular type changes from open-source projects to improve built-in IDE refactoring capabilities.

The instructions for downloading, installing, and using the plugin are available online.\footnote{\toolname: \url{https://type-change.github.io/}}
\section{\toolname}\label{sec:implementation}

\begin{figure}
    \centering
    \includegraphics[width=\columnwidth]{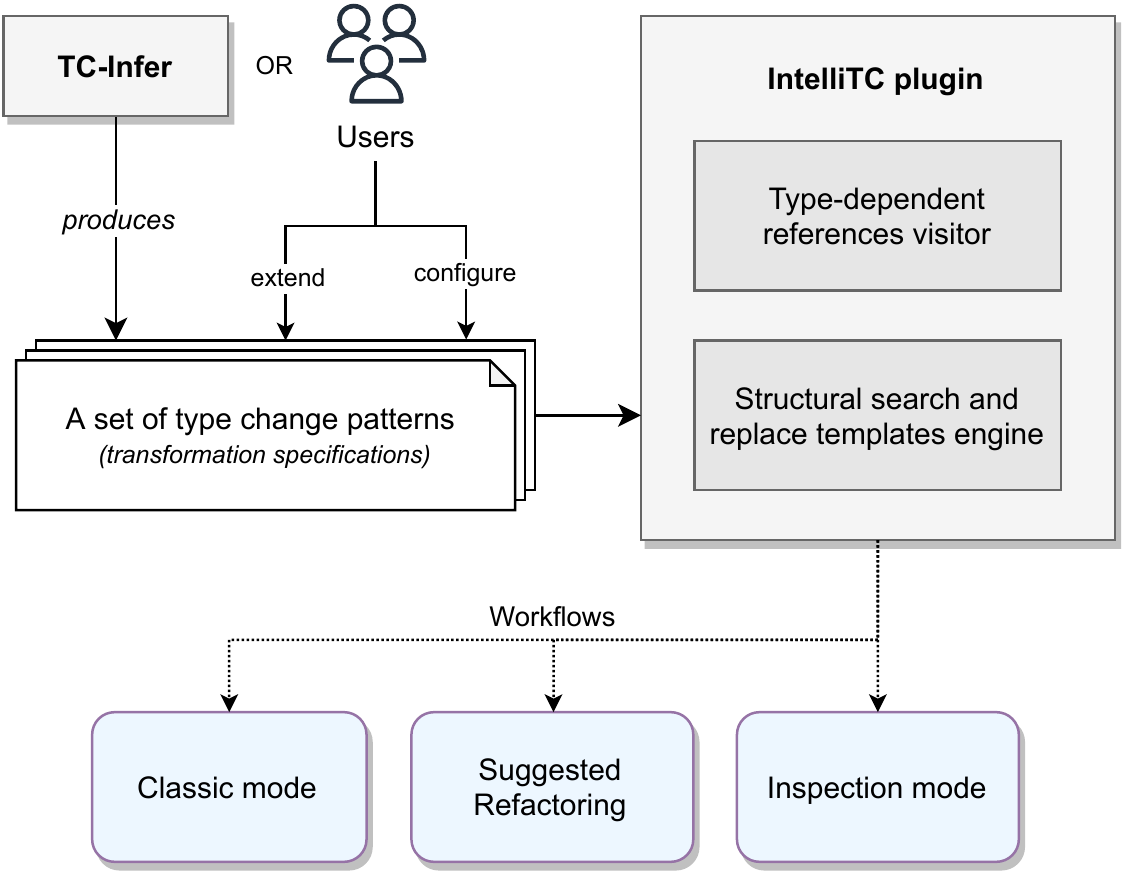}
    \caption{The pipeline behind \toolname.}
    \label{fig:pipeline}
\end{figure}

\subsection{Overview}

The high-level overview of the plugin is presented in ~\Cref{fig:pipeline}.
The plugin allows developers to automatically perform a set of pre-configured type changes (inferred by \tcinfer) and to configure any custom type changes by themselves. 
\toolname's core components are implemented with the use of the IntelliJ Platform SDK.\footnote{The IntelliJ Platform: \url{https://plugins.jetbrains.com/docs/intellij/intellij-platform.html}}
Using the plugin, the developer can follow 3 different workflows.
First, \toolname is manually invoked at a variable or a method (the \textit{root} element of the transformation) to automatically perform the desired type change.
Second, \toolname tracks the developer's activity in the code editor to understand their intent and appropriately \textit{suggests} a type change refactoring.
Third, \toolname recommends certain type changes by \textit{inspecting} the code.

\begin{figure}[t]
    \centering
    \begin{lstlisting}[language=json,numbers=none]
  {
    "From": "java.io.File",
    "To": "java.nio.file.Path",
    "ID": 1, 
    "Priority": 2, 
    "Mode": "Suggested Refactoring",
    "Rules": [
      {
        "Before": "new File($1$, $2$)",
        "After": "$1$.resolve($2$)"
      },
      {
        "Before": "$1$.exists()",
        "After": "Files.exists($1$)",
      },
      {
        "Before": "$1$.toPath()",
        "After": "$1$"
      },
      ...
    ]
  }
    \end{lstlisting}
    \caption{
        A fragment of the ``\smcode{File}$\rightarrow$\smcode{Path}'' type change pattern. In the ``Rules" section, the template variable \smcode{\$1\$} is responsible for matching the \textit{root} element of the type change, whereas the template variable \smcode{\$2\$} is used for matching any other AST expression node greedily, except the root.
    }
    \label{fig:tc}
\end{figure}

\begin{figure*}[ht]
    \centering
    \includegraphics[width=\textwidth]{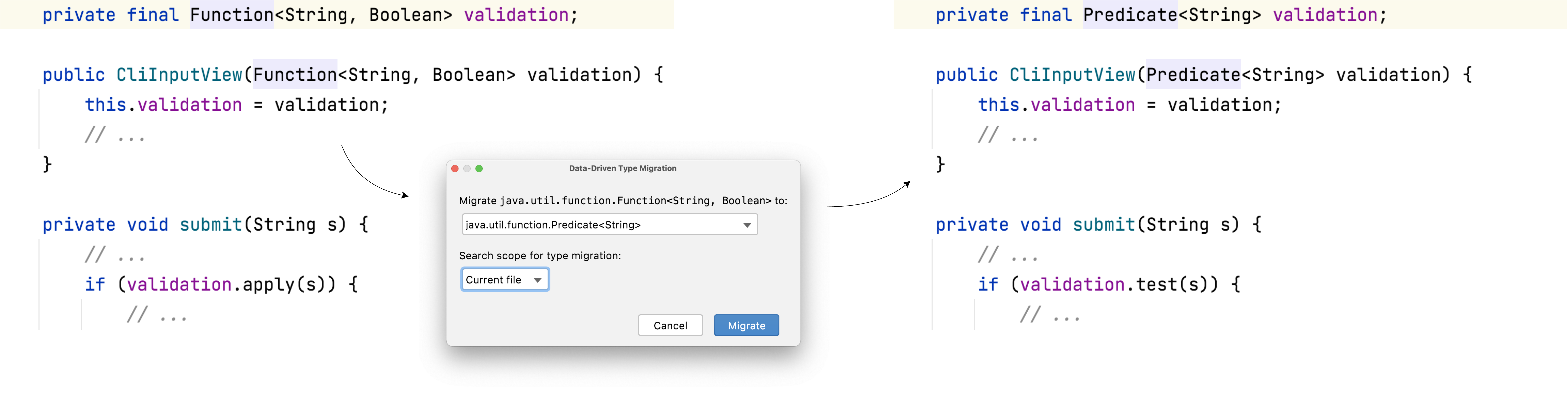}
    \caption{An example usage of the \textsc{IntelliTC} plugin applied upon a code snippet from the \textsc{Apache Flink} project. We submitted a pull request with 9 type changes (including this) aimed at eliminating such misuse, and it was accepted by the maintainers~\cite{ketkar2022tcinfer}.
    }
    \label{fig:usage}
\end{figure*}

\subsection{Transformation Specifications}

Developers can edit existing rules, add new rules to existing type change patterns, or even add new type change patterns by updating the \textit{transformation specifications} JSON exposed by \toolname via the \textit{Settings} tab of IntelliJ IDEA, as shown in \Cref{fig:tc}. 
The schema followed by this JSON is analogous to the output \textit{transformation specifications} produced by \tcinfer, where each type change pattern (\ie \code{(source-type, target-type)}) is associated with a set of rewrite rules over Java statements (and expressions). 
One of these associated rewrite rules will be applied to adapt each reference when the type change is performed.

In addition to this, \toolname also allows users to rank the type changes as they will appear in the UI (\textit{Priority} in \Cref{fig:tc}), and specify how to surface the type change suggestion (\textit{Classic mode}, \textit{Inspection mode}, \textit{Suggested Refactoring}~---~described in \Cref{modalities}). 
By default, all type change patterns specified in this JSON can be applied only by manually invoking the plugin and are not automatically surfaced.

In \toolname, we include a set of manually vetted rewrite rules for 59 popular and useful type change patterns that we constructed as a part of the evaluation of \tcinfer (described in our accompanying paper's RQ 3~\cite{ketkar2022tcinfer}). 
There, we investigated a corpus of 129 large, mature, and diverse Java projects and identified 40,865 commits where type changes were performed. 
We then analysed these commits and found 605 \textit{popular} type change patterns that were performed in more than one unique project. 
We applied \tcinfer on these commits to infer the associated rewrite rules for these 605 type change patterns. 
From these patterns, we manually selected 59 type change patterns to be used as an \textit{initial} input for \toolname, based on their popularity and relevance for the end-use developers (we only considered the type changes between built-in JDK types). 
We also automatically calculated \textit{Priority} based on the number of commits where type change was performed, and manually labelled the suggestion level for these 59 type change patterns (field \textit{Mode} in \Cref{fig:tc}).

\subsection{Implementation}

To handle the search of type-dependent references in Java code, we decided to leverage the capabilities of the existing IntelliJ's Type Migration framework~\cite{typeMigration}. 
\toolname reuses its core visitor components to search for all the candidate source code locations for update (by inter-procedural analysis across type constraints) and then performs the actual update by executing IntelliJ's Structural-Search-and-Replace (SSR) templates~\cite{SSR}.
Such templates allow matching the code fragments in a regex-like manner, considering their tree structure and also leaving \textit{holes} (like \smcode{\$1\$} and \smcode{\$2\$} in \Cref{fig:tc}) for tree nodes, which is especially useful when matching identifiers.
Using SSR lowers the barrier of entry for new users of the tool, since SSR is a well-documented part of the IntelliJ Platform~\cite{SSR}.

\toolname lets users define a \textit{scope} for reference search, including \textit{Local} scope (the enclosing method of the identifier which type is being changed), \textit{Current File/Class} scope, and \textit{Global/Project} scope.
The plugin also supports \textit{undoing} the type change.

Under the hood, \toolname runs the visitors provided by IntelliJ's Type Migration framework, and updates the type-dependent references with the rewrite rules contained in the chosen type change pattern.
For each reference, the final rewrite rule is chosen by the largest number of matched code tokens between the reference itself (or its parent in the AST) and the before-part of some rewrite rule.

\subsection{Modalities}\label{modalities}
\toolname operates in three different modes:

\subsubsection{Classic mode}
In this mode, the plugin operates as a general \textit{code intention},\footnote{Intention Actions: \url{https://www.jetbrains.com/help/idea/intention-actions.html}} providing the developers with the ability to apply a type change refactoring only when it is invoked directly from the context menu of some type element in the code.
For instance, as it is shown in \Cref{fig:usage}, developers can invoke intention action for the type of the field \smcode{validation} (which defines a \textit{root} element here), aiming to change it from \smcode{Function<String, Boolean>} to \smcode{Predicate<String>}.
They can specify the type change pattern along with a search scope in the shown dialog box.
If \toolname cannot update any references of the \textit{root} element, it will show the \textit{Tool Window} with the failed usages.
This allows developers to fix the problems manually or use built-in quick-fixes.

\subsubsection{Suggested Refactoring}
Previous researchers~\cite{Ge:ICSE:RefactoringSteps,witchdoctor} observed that \textit{discoverability} and \textit{late awareness} led to the underuse of refactoring tools. 
To counter this problem, we leverage IntelliJ's \textit{Suggested Refactoring} mechanism.
In this mode, a corresponding type change refactoring is suggested by \toolname when the user manually changes the type of some element in Java code. 
The plugin tracks such changes in the document model and renders the button in the left-side panel of the code editor, allowing the user to click it, configure the necessary search scope, and run the refactoring.

\label{isomorphic}
Note that not all type changes are applicable in each context (\eg \smcode{String} to \smcode{Pattern}, or \smcode{String} to \smcode{Path}), and receiving spurious suggestions from the IDE could confuse developers.
Thus, we decided to make Suggested Refactoring available only for \textit{isomorphic} types (that are interchangeable), like \smcode{File} and \smcode{Path}, or \smcode{Date} and \smcode{LocalDate}.
Currently, the user needs to manually label the isomorphic types (via the \smcode{"Mode"} field in \Cref{fig:tc}) in the input JSON. 
We believe that automatically detecting isomorphic type changes and suggesting them is a challenging yet promising direction for future work.

\subsubsection{Inspection mode}
The last but not least is the mode in which \toolname runs as a \textit{code inspection}.\footnote{Code Inspections: \url{https://www.jetbrains.com/help/idea/code-inspection.html}}
Code inspections are performed automatically by the IntelliJ Platform's engine in the background, and are useful for detecting possible problems with the code.
\toolname's \textit{Inspection mode} leverages this interface to provide quick fixes involving type changes. 
Currently, \toolname promotes the clean-code recommendations from Effective Java~\cite{bloch2008effective},
like eliminating misuses of Java 8 functional interfaces (\eg \smcode{Function<T, Boolean>$\rightarrow$Predicate<T>}, see \Cref{fig:usage}). 
\toolname highlights such occurrences of misused types in the program and provides the appropriate intention actions to replace them.
\section{Evaluation}\label{sec:evaluation}

\begin{figure}
    \centering
    \includegraphics[width=\columnwidth]{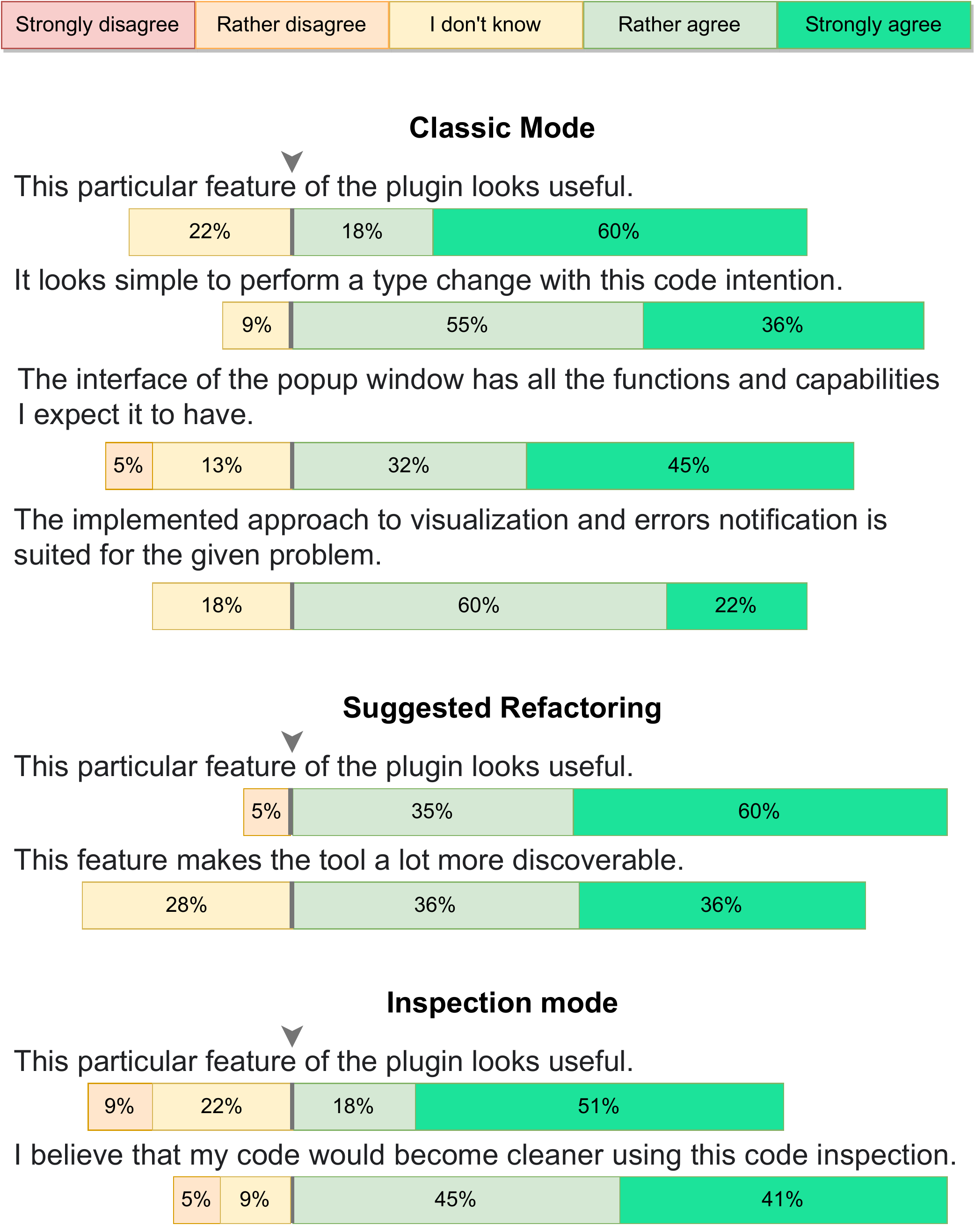}
    \caption{The results of the conducted preliminary survey.}
    \label{table:results}
    \vspace{-0.1em}
\end{figure}

Our previous work~\cite{ketkar2022tcinfer} provides an in-depth empirical evaluation for \tcinfer and shows the merit of using a data-driven approach. 
For this demo paper, we complement the previous thorough evaluation with a survey to determine how developers apply type changes in their everyday work, and asked them to evaluate the usefulness of the \toolname for different usage modalities.
We intentionally designed the study such that participants are able to see a short and feature-focused demo for each mode of the plugin, instead of forcing them to install and run it.
The previous research~\cite{kery2020mage, davidoff2007rapidly} has shown that such method could be helpful to evaluate the core concepts and ideas behind the tool, whereas high-fidelity prototypes might distract survey participants with a lot of implementation details.

We have questioned 22 qualified software developers with 2 to 5 years of professional experience on average (in particular, four of them had been working with Java for 10+ years).
All the respondents confirmed that they frequently use IntelliJ IDEA's code intentions, inspections, and automated refactoring features.
Even so, almost half of the participants had also mentioned that in everyday practice, their automated refactoring experience comes down to some simple scenarios, such as renames.
Twelve people stated their awareness of existing Type Migration refactoring, but only \textit{3} of them had actually used it.
It should also be noted that our evaluation is limited by considering the developers who \textit{already} use refactoring tools, thus assessment of people who have no experience with such tools could affect the final results.

We asked the respondents to evaluate the usefulness and compare the three different modes of providing type change refactoring (see~\Cref{sec:implementation}) in the plugin.
We have used Likert-type~\cite{boone2012analyzing} questions and followed the best-practices from existing usability studies~\cite{barnum2010usability}. 
\Cref{table:results} shows the results. 
Notice that the participants have highly positive attitudes towards the usefulness and the interfaces of \toolname.
We also asked them to compare the existing plugin modes, and \textit{Suggested Refactoring} was chosen as the most useful by 80\% of developers.
However, approximately half of all participants expressed their convictions to receive such suggestions for \textit{isomorphic} types only (see ~\Cref{isomorphic}).

The developers have also left constructive feedback on how they envision improving \toolname.
They expressed their need to see a preview of the refactoring, and also to receive additional motivation for the recommended type changes in the \textit{Inspection} mode.
We plan to implement these features as a part of our future work.

We are also grateful to receive such inspiring positive reviews:

\begin{quote}
    ``It really does save a lot of time and energy, as a String can be used in lots of places in code. I usually coped with changing Strings to Patterns manually, which was really boring and time consuming.''
\end{quote}

\begin{quote}
    ``If I knew that this exists, I would definitely use it.''
\end{quote}

\section{Related Work}

Previous researchers and tool-builders have proposed several frameworks~\cite{balaban2005refactoring, ketkar2019type, wright2020incremental} to address the problem of applying custom type changes in source code.
These approaches require API transformation specifications to be manually encoded in the corresponding \textit{domain specific language}.
However, Kim et al.~\cite{scripting:kim:batory} have shown that encoding refactorings via DSL might be overwhelming, and such activity has a steep learning curve. 
To overcome this barrier of using type migration tools, we proposed \tcinfer in our accompanying paper~\cite{ketkar2022tcinfer} to automatically infer rewrite rules for common type changes, while \toolname focuses on automating them in the IDE in a developer-friendly manner.

From the standpoint of production-ready tools, there is no actual support for \textit{custom} type change refactoring or class library migration in the current IDEs.  
IntelliJ IDEA~\cite{IDEA}, being the most popular IDE for Java developers in 2021~\cite{java2021eco}, provides \textit{type migration} refactoring only for a small set of pre-defined types, however it does not support custom API rewrite rules. 
To the best of our knowledge, other popular IDEs for Java (such as Eclipse~\cite{eclipse}, Visual Studio Code~\cite{vscode}, NetBeans~\cite{netbeans}) do not provide any functionality for performing type changes.
\toolname allows developers to define custom type change rules using the \textit{Structural Search and Replace} templates and automates these for the developer in the IDE.

Previously researchers have also developed tools to perform library migrations or library updates at the source code level:
\begin{enumerate*}
    \item ~\citet{xi2019migrating} proposed a tool called \textsc{DAAMT} which automates the migration of deprecated Java APIs based on its replacements from the documentation,
    \item ~\citet{lamothe2020a3} proposed a way to migrate deprecated Android APIs by learning from code examples,
    \item ~\citet{collie2020m3} employed a probabilistic program synthesis to tackle cases when there is no prior knowledge of the target library API usage. 
\end{enumerate*}
In contrast to these works, this paper largely focuses on the \textit{user experience} aspect of the tool, and addresses the problem of effectively performing a \textit{type change} refactoring in an IDE.

\section{Conclusion}\label{sec:conclusion}

In this paper, we present \toolname~---~an IntelliJ IDEA plugin for automating type changes.
Our approach uses custom API transformation specifications, which are automatically mined and inferred from the history of code changes by \tcinfer~\cite{ketkar2022tcinfer}, or added and tweaked by the users themselves.
We have presented three ways for providing type change refactoring opportunity to the developer from the standpoint of user experience and conducted a preliminary evaluation of its potential usefulness with 22 software developers.
\toolname was warmly received by the participants of the study, and we plan to continue improving it based on their feedback. 

\section{Acknowledgements}
We would like to thank Zarina Kurbatova for her help with the technical details of working with the IntelliJ Platform, Souti Chattopadhyay for helping us design the evaluation, Yaroslav Golubev and the reviewers for their constructive feedback to improve the work.
We also thank our survey participants for their insightful answers.

\bibliographystyle{ACM-Reference-Format}
\balance
\bibliography{bibliography}

%%% -*-BibTeX-*-
%%% Do NOT edit. File created by BibTeX with style
%%% ACM-Reference-Format-Journals [18-Jan-2012].

\begin{thebibliography}{26}

%%% ====================================================================
%%% NOTE TO THE USER: you can override these defaults by providing
%%% customized versions of any of these macros before the \bibliography
%%% command.  Each of them MUST provide its own final punctuation,
%%% except for \shownote{}, \showDOI{}, and \showURL{}.  The latter two
%%% do not use final punctuation, in order to avoid confusing it with
%%% the Web address.
%%%
%%% To suppress output of a particular field, define its macro to expand
%%% to an empty string, or better, \unskip, like this:
%%%
%%% \newcommand{\showDOI}[1]{\unskip}   % LaTeX syntax
%%%
%%% \def \showDOI #1{\unskip}           % plain TeX syntax
%%%
%%% ====================================================================

\ifx \showCODEN    \undefined \def \showCODEN     #1{\unskip}     \fi
\ifx \showDOI      \undefined \def \showDOI       #1{#1}\fi
\ifx \showISBNx    \undefined \def \showISBNx     #1{\unskip}     \fi
\ifx \showISBNxiii \undefined \def \showISBNxiii  #1{\unskip}     \fi
\ifx \showISSN     \undefined \def \showISSN      #1{\unskip}     \fi
\ifx \showLCCN     \undefined \def \showLCCN      #1{\unskip}     \fi
\ifx \shownote     \undefined \def \shownote      #1{#1}          \fi
\ifx \showarticletitle \undefined \def \showarticletitle #1{#1}   \fi
\ifx \showURL      \undefined \def \showURL       {\relax}        \fi
% The following commands are used for tagged output and should be
% invisible to TeX
\providecommand\bibfield[2]{#2}
\providecommand\bibinfo[2]{#2}
\providecommand\natexlab[1]{#1}
\providecommand\showeprint[2][]{arXiv:#2}

\bibitem[Apache(1997)]%
        {netbeans}
\bibfield{author}{\bibinfo{person}{Apache}.} \bibinfo{year}{1997}\natexlab{}.
\newblock \bibinfo{booktitle}{\emph{NetBeans}}.
\newblock
\urldef\tempurl%
\url{https://netbeans.apache.org/}
\showURL{%
\tempurl}


\bibitem[Balaban et~al\mbox{.}(2005)]%
        {balaban2005refactoring}
\bibfield{author}{\bibinfo{person}{Ittai Balaban}, \bibinfo{person}{Frank Tip},
  {and} \bibinfo{person}{Robert Fuhrer}.} \bibinfo{year}{2005}\natexlab{}.
\newblock \showarticletitle{Refactoring support for class library migration}.
\newblock \bibinfo{journal}{\emph{ACM SIGPLAN Notices}} \bibinfo{volume}{40},
  \bibinfo{number}{10} (\bibinfo{year}{2005}), \bibinfo{pages}{265--279}.
\newblock


\bibitem[Barnum(2010)]%
        {barnum2010usability}
\bibfield{author}{\bibinfo{person}{Carol~M Barnum}.}
  \bibinfo{year}{2010}\natexlab{}.
\newblock \bibinfo{booktitle}{\emph{Usability testing essentials}}.
\newblock \bibinfo{publisher}{Elsevier}.
\newblock


\bibitem[Bloch(2008)]%
        {bloch2008effective}
\bibfield{author}{\bibinfo{person}{Joshua Bloch}.}
  \bibinfo{year}{2008}\natexlab{}.
\newblock \bibinfo{booktitle}{\emph{Effective java}}.
\newblock \bibinfo{publisher}{Addison-Wesley Professional}.
\newblock


\bibitem[Boone and Boone(2012)]%
        {boone2012analyzing}
\bibfield{author}{\bibinfo{person}{Harry~N Boone} {and}
  \bibinfo{person}{Deborah~A Boone}.} \bibinfo{year}{2012}\natexlab{}.
\newblock \showarticletitle{Analyzing likert data}.
\newblock \bibinfo{journal}{\emph{Journal of extension}} \bibinfo{volume}{50},
  \bibinfo{number}{2} (\bibinfo{year}{2012}), \bibinfo{pages}{1--5}.
\newblock


\bibitem[Collie et~al\mbox{.}(2020)]%
        {collie2020m3}
\bibfield{author}{\bibinfo{person}{Bruce Collie}, \bibinfo{person}{Philip
  Ginsbach}, \bibinfo{person}{Jackson Woodruff}, \bibinfo{person}{Ajitha
  Rajan}, {and} \bibinfo{person}{Michael~FP O'Boyle}.}
  \bibinfo{year}{2020}\natexlab{}.
\newblock \showarticletitle{M3: Semantic api migrations}. In
  \bibinfo{booktitle}{\emph{2020 35th IEEE/ACM International Conference on
  Automated Software Engineering (ASE)}}. IEEE, \bibinfo{pages}{90--102}.
\newblock


\bibitem[Davidoff et~al\mbox{.}(2007)]%
        {davidoff2007rapidly}
\bibfield{author}{\bibinfo{person}{Scott Davidoff}, \bibinfo{person}{Min~Kyung
  Lee}, \bibinfo{person}{Anind~K Dey}, {and} \bibinfo{person}{John Zimmerman}.}
  \bibinfo{year}{2007}\natexlab{}.
\newblock \showarticletitle{Rapidly exploring application design through speed
  dating}. In \bibinfo{booktitle}{\emph{International Conference on Ubiquitous
  Computing}}. Springer, \bibinfo{pages}{429--446}.
\newblock


\bibitem[{Foster} et~al\mbox{.}(2012)]%
        {witchdoctor}
\bibfield{author}{\bibinfo{person}{S.~R. {Foster}}, \bibinfo{person}{W.~G.
  {Griswold}}, {and} \bibinfo{person}{S. {Lerner}}.}
  \bibinfo{year}{2012}\natexlab{}.
\newblock \showarticletitle{WitchDoctor: IDE support for real-time
  auto-completion of refactorings}. In \bibinfo{booktitle}{\emph{ICSE}}.
  \bibinfo{pages}{222--232}.
\newblock
\urldef\tempurl%
\url{https://doi.org/10.1109/ICSE.2012.6227191}
\showDOI{\tempurl}


\bibitem[Foundation(2001)]%
        {eclipse}
\bibfield{author}{\bibinfo{person}{Eclipse Foundation}.}
  \bibinfo{year}{2001}\natexlab{}.
\newblock \bibinfo{booktitle}{\emph{Eclipse}}.
\newblock
\urldef\tempurl%
\url{https://www.eclipse.org/}
\showURL{%
\tempurl}


\bibitem[Ge et~al\mbox{.}(2012)]%
        {Ge:ICSE:RefactoringSteps}
\bibfield{author}{\bibinfo{person}{Xi Ge}, \bibinfo{person}{Quinton~L. DuBose},
  {and} \bibinfo{person}{Emerson Murphy-Hill}.}
  \bibinfo{year}{2012}\natexlab{}.
\newblock \showarticletitle{Reconciling Manual and Automatic Refactoring}. In
  \bibinfo{booktitle}{\emph{ICSE}} (Zurich, Switzerland)
  \emph{(\bibinfo{series}{ICSE '12})}. \bibinfo{publisher}{IEEE Press},
  \bibinfo{pages}{211–221}.
\newblock
\showISBNx{9781467310673}


\bibitem[Google(2011)]%
        {error-prone}
\bibfield{author}{\bibinfo{person}{Google}.} \bibinfo{year}{2011}\natexlab{}.
\newblock \bibinfo{booktitle}{\emph{Error Prone}}.
\newblock
\urldef\tempurl%
\url{https://github.com/google/error-prone}
\showURL{%
\tempurl}


\bibitem[JetBrains(2021a)]%
        {IDEA}
\bibfield{author}{\bibinfo{person}{JetBrains}.}
  \bibinfo{year}{2021}\natexlab{a}.
\newblock \bibinfo{booktitle}{\emph{IntelliJ IDEA official website}}.
\newblock
\urldef\tempurl%
\url{https://www.jetbrains.com/idea/}
\showURL{%
\tempurl}


\bibitem[JetBrains(2021b)]%
        {SSR}
\bibfield{author}{\bibinfo{person}{JetBrains}.}
  \bibinfo{year}{2021}\natexlab{b}.
\newblock \bibinfo{booktitle}{\emph{{IntelliJ's Structural Search and
  Replace}}}.
\newblock
\urldef\tempurl%
\url{https://www.jetbrains.com/help/idea/structural-search-and-replace.html}
\showURL{%
\tempurl}


\bibitem[JetBrains(2021c)]%
        {typeMigration}
\bibfield{author}{\bibinfo{person}{JetBrains}.}
  \bibinfo{year}{2021}\natexlab{c}.
\newblock \bibinfo{booktitle}{\emph{{IntelliJ's Type Migration framework}}}.
\newblock
\urldef\tempurl%
\url{https://www.jetbrains.com/help/idea/type-migration.html}
\showURL{%
\tempurl}


\bibitem[JetBrains(2021d)]%
        {java2021eco}
\bibfield{author}{\bibinfo{person}{JetBrains}.}
  \bibinfo{year}{2021}\natexlab{d}.
\newblock \bibinfo{booktitle}{\emph{The State of Java Developer Ecosystem}}.
\newblock
\urldef\tempurl%
\url{https://www.jetbrains.com/lp/devecosystem-2021/java/}
\showURL{%
\tempurl}


\bibitem[Johnson(2001)]%
        {johnson2001you}
\bibfield{author}{\bibinfo{person}{Philip~M Johnson}.}
  \bibinfo{year}{2001}\natexlab{}.
\newblock \showarticletitle{You can’t even ask them to push a button: Toward
  ubiquitous, developer-centric, empirical software engineering}. In
  \bibinfo{booktitle}{\emph{The NSF Workshop for New Visions for Software
  Design and Productivity: Research and Applications}}. Citeseer.
\newblock


\bibitem[Kery et~al\mbox{.}(2020)]%
        {kery2020mage}
\bibfield{author}{\bibinfo{person}{Mary~Beth Kery}, \bibinfo{person}{Donghao
  Ren}, \bibinfo{person}{Fred Hohman}, \bibinfo{person}{Dominik Moritz},
  \bibinfo{person}{Kanit Wongsuphasawat}, {and} \bibinfo{person}{Kayur Patel}.}
  \bibinfo{year}{2020}\natexlab{}.
\newblock \showarticletitle{mage: Fluid Moves Between Code and Graphical Work
  in Computational Notebooks}. In \bibinfo{booktitle}{\emph{Proceedings of the
  33rd Annual ACM Symposium on User Interface Software and Technology}}.
  \bibinfo{pages}{140--151}.
\newblock


\bibitem[Ketkar et~al\mbox{.}(2019)]%
        {ketkar2019type}
\bibfield{author}{\bibinfo{person}{Ameya Ketkar}, \bibinfo{person}{Ali Mesbah},
  \bibinfo{person}{Davood Mazinanian}, \bibinfo{person}{Danny Dig}, {and}
  \bibinfo{person}{Edward Aftandilian}.} \bibinfo{year}{2019}\natexlab{}.
\newblock \showarticletitle{Type migration in ultra-large-scale codebases}. In
  \bibinfo{booktitle}{\emph{ICSE}}. IEEE, \bibinfo{pages}{1142--1153}.
\newblock


\bibitem[Ketkar et~al\mbox{.}(2022)]%
        {ketkar2022tcinfer}
\bibfield{author}{\bibinfo{person}{Ameya Ketkar}, \bibinfo{person}{Oleg
  Smirnov}, \bibinfo{person}{Nikolaos Tsantalis}, \bibinfo{person}{Danny Dig},
  {and} \bibinfo{person}{Timofey Bryksin}.} \bibinfo{year}{2022}\natexlab{}.
\newblock \showarticletitle{Inferring and Applying Type Changes}. In
  \bibinfo{booktitle}{\emph{44th International Conference on Software
  Engineering (ICSE '22)}} (Pittsburgh, United States)
  \emph{(\bibinfo{series}{ICSE ’22})}. ACM.
\newblock
\urldef\tempurl%
\url{https://doi.org/10.1145/3510003.3510115}
\showDOI{\tempurl}


\bibitem[Ketkar et~al\mbox{.}(2020)]%
        {ketkar2020understanding}
\bibfield{author}{\bibinfo{person}{Ameya Ketkar}, \bibinfo{person}{Nikolaos
  Tsantalis}, {and} \bibinfo{person}{Danny Dig}.}
  \bibinfo{year}{2020}\natexlab{}.
\newblock \showarticletitle{Understanding type changes in java}. In
  \bibinfo{booktitle}{\emph{FSE}}. \bibinfo{pages}{629--641}.
\newblock


\bibitem[Kim et~al\mbox{.}(2015)]%
        {scripting:kim:batory}
\bibfield{author}{\bibinfo{person}{Jongwook Kim}, \bibinfo{person}{Don Batory},
  {and} \bibinfo{person}{Danny Dig}.} \bibinfo{year}{2015}\natexlab{}.
\newblock \showarticletitle{Scripting parametric refactorings in Java to
  retrofit design patterns}. In \bibinfo{booktitle}{\emph{2015 IEEE
  International Conference on Software Maintenance and Evolution (ICSME)}}.
  \bibinfo{pages}{211--220}.
\newblock
\urldef\tempurl%
\url{https://doi.org/10.1109/ICSM.2015.7332467}
\showDOI{\tempurl}


\bibitem[Lamothe et~al\mbox{.}(2020)]%
        {lamothe2020a3}
\bibfield{author}{\bibinfo{person}{Maxime Lamothe}, \bibinfo{person}{Weiyi
  Shang}, {and} \bibinfo{person}{Tse-Hsun~Peter Chen}.}
  \bibinfo{year}{2020}\natexlab{}.
\newblock \showarticletitle{A3: Assisting android api migrations using code
  examples}.
\newblock \bibinfo{journal}{\emph{IEEE Transactions on Software Engineering}}
  (\bibinfo{year}{2020}).
\newblock


\bibitem[LLVM(2007)]%
        {clang}
\bibfield{author}{\bibinfo{person}{LLVM}.} \bibinfo{year}{2007}\natexlab{}.
\newblock \bibinfo{booktitle}{\emph{Clang-tidy}}.
\newblock
\urldef\tempurl%
\url{https://clang.llvm.org/extra/clang-tidy/}
\showURL{%
\tempurl}


\bibitem[Microsoft(2015)]%
        {vscode}
\bibfield{author}{\bibinfo{person}{Microsoft}.}
  \bibinfo{year}{2015}\natexlab{}.
\newblock \bibinfo{booktitle}{\emph{Visual Studio Code}}.
\newblock
\urldef\tempurl%
\url{https://code.visualstudio.com/}
\showURL{%
\tempurl}


\bibitem[Wright(2020)]%
        {wright2020incremental}
\bibfield{author}{\bibinfo{person}{Hyrum~K Wright}.}
  \bibinfo{year}{2020}\natexlab{}.
\newblock \showarticletitle{Incremental type migration using type algebra}. In
  \bibinfo{booktitle}{\emph{ICSME}}. IEEE, \bibinfo{pages}{756--765}.
\newblock


\bibitem[Xi et~al\mbox{.}(2019)]%
        {xi2019migrating}
\bibfield{author}{\bibinfo{person}{Yaoguo Xi}, \bibinfo{person}{Liwei Shen},
  \bibinfo{person}{Yukun Gui}, {and} \bibinfo{person}{Wenyun Zhao}.}
  \bibinfo{year}{2019}\natexlab{}.
\newblock \showarticletitle{Migrating deprecated api to documented replacement:
  Patterns and tool}. In \bibinfo{booktitle}{\emph{Proceedings of the 11th
  Asia-Pacific Symposium on Internetware}}. \bibinfo{pages}{1--10}.
\newblock


\end{thebibliography}

\end{document}